\documentclass[usenatbib]{mn2e}
\usepackage{epsfig}
\usepackage{amsmath,amssymb}
\newcommand{\hMsol}{{\,h^{-1}\rm M}_\odot}
\newcommand{\hMpc}{{\,h^{-1}\rm Mpc}}

\newcommand{\CF}{{\cal F}}
\newcommand{\paverage}[1]{\left\langle #1 \right\rangle_{\rm P}}
\renewcommand{\vec}[1]{{\mathbf #1}}
\title{Imprints of mass accretion on properties of galaxy clusters}
\author[Faltenbacher et al.]
{\parbox[t]\textwidth{Andreas Faltenbacher$^{1,5}$, Brandon Allgood$^2$, 
Stefan Gottl\"ober$^3$, Gustavo Yepes$^4$
and Yehuda Hoffman$^5$}
\vspace*{6pt} \\
$^1$UCO/Lick Observatory,
University of California at Santa Cruz, 
1156 High Street, Santa Cruz, CA 95064, USA
\\
$^2$Physics Dept., University of California at Santa Cruz,
1156 High Street, Santa Cruz, CA 95064, USA
\\
$^3$Astrophysikalisches Institut Potsdam,
An der Sternwarte 16, 14482 Potsdam, Germany
\\
$^4$Grupo de Astrof\'{\i}sica, 
Universidad Aut\'onoma de Madrid,
Madrid E-280049, Spain 
\\
$^5$Racah Institute of Physics, 
Hebrew University, 
Jerusalem 91904, Israel 
\\
}
\date{\today}
\begin{document}
\maketitle
\begin{abstract}
A large scale SPH+N-body simulation (GADGET) of the concordance
$\Lambda$CDM universe is used to investigate orientation and angular
momentum of galaxy clusters at $z=0$ in connection with their recent
accretion histories. The basic cluster sample comprises the 3000 most
massive  friends-of-friends halos found in the $500\hMpc$ simulation
box. Two disjoint sub-samples are constructed, using the mass
ratio of the two most massive progenitors at $z=0.5$ $m_2/m_1$
($m_1\geq m_2$), namely a {\it recent major merger} sample and a {\it
steady accretion mode} sample. The mass of clusters in the merger
sample is on average $\sim43\%$ larger than the mass of the two
progenitors ($m_1 + m_2$), whereas in the steady accretion mode sample
a smaller increase of $\sim25\%$ is found. The separation vector 
connecting the two most massive progenitor halos at $z=0.5$ is
strongly correlated with the orientation of the cluster at $z=0$. The 
angular momentum of the clusters in the recent major merger sample
tends to be parallel to orbital angular momentum of the two
progenitors, whereas the angular momentum of the steady accretion mode
sample is mainly determined by the angular momentum of the most massive
progenitor. The long range correlations for the major and the minor
principal axes of cluster pairs extend to distances of
$\sim100\hMpc$. Weak angular momentum correlations are found for
distances $\lesssim 20\hMpc$. Within these ranges the major axes tend
to be aligned with the connecting line of the cluster pairs whereas minor
axes and angular momenta tend to be perpendicular to this line. A
separate analysis of the two sub-samples reveals that the long range
correlations are independent of the mass accretion mode. Thus
orientation and angular momentum of galaxy clusters is mainly
determined by the accretion along the filaments independently of the
particular accretion mode.       
\end{abstract}
\begin{keywords} 
methods: numerical -- methods: statistical --  galaxies: clusters: general
-- large-scale structure of the Universe
\end{keywords}
\section{Introduction}
Using a sample of 44 Abell clusters \cite{binggeli:shape} showed that
clusters of galaxies are highly eccentric. \cite{mcmillan_etal1989}
measured the X-ray contours of 49 Abell clusters and found that
most clusters are clearly flattened. By means of N-body simulations
\cite{warren_etal1992} (see also \citealt{dubinski_carlberg1991,
cole_lacey1996}) demonstrated that dark matter halos have triaxial
shapes and tend to be prolate. Furthermore they showed that the
prevailing shape is supported by the anisotropic 
velocity dispersion (see also \cite{tormen1997}). Thus the
shape of the halo mirrors its dynamical state and is
likely to be a long-term feature if no major disruptions take
place. Subsequently the {\it orientation} of clusters will be
identified with the major axis of its mass distribution.   

The {\it angular momentum} of galaxy clusters is easily accessible in
numerical simulations. The orientation of the angular momentum may
record information about the mass accretion history and/or tidal
interactions with the surrounding large scale structure. According to 
\cite{doroshkevich:origin}, \cite{peebles:LSS}, and
\cite{white:angular} the primary angular momentum of bound objects is 
due to tidal interaction between the elongated proto-structures after
decoupling from cosmic expansion and before turn-around.  More recent
studies find that the angular momentum of dark matter halos is later
modified by the merging history of their building blocks, see
\cite{vitvitska:origin} and \cite{porciani:testingI,porciani:testingII}.     

The {\it alignment} of galaxy cluster orientations is controversial
since \cite{binggeli:shape} published his inaugural study. He reported
that galaxy clusters are oriented relative to neighbours at least for
separations smaller than $15\hMpc$. Moreover, he found anisotropies in
the cluster distribution on scales up to $50\hMpc$. 
The first investigations, using X-ray contours, led
to no \citep{ulmer:major} or only weak \citep{rhee:x+opt} significance
of alignment effects. However, \cite{chambers:nearest} exploring X-ray
position angles of 45 clusters found significant alignment of clusters'
orientations with the connecting line to the next neighbour for
separations of $\leq 20\hMpc$. Analysing a large set of optical data
of 637 Abell clusters \citet{plionis:up150} found highly significant
alignment effects on scales below $10\hMpc$ that become weaker but
extend up to $150\hMpc$. 

In the literature two slightly different scenarios are proposed
to explain the alignment of the orientation of
clusters. \cite{binney:prolat} suggested that tidal interactions of
evolving proto-cluster systems may lead to the growth of anisotropies
of clusters and to relative orientation effects. However, \cite{west:merging}
found that clusters grow by accretion and merging of surrounding
matter that falls into the deep cluster potential wells along
sheet-like and filamentary high density regions.
Using numerical CDM simulations \cite{vanHaarlem:merging} 
demonstrate that clusters are elongated along the incoming direction
of the last major merger.  Further
\cite{west:principal} show that Virgo's brightest elliptical
galaxies have a remarkably collinear arrangement in three dimensions, 
which is aligned with the filamentary structure connecting Virgo and
the rich cluster Abell 1367. In a statistical investigation of 303
Abell clusters \cite{plionis:galali} confirm the alignments of galaxy
members with their parent cluster orientation as well as with the
large-scale environment within which the clusters are embedded. 
Therefore, the cluster formation is tightly connected with the
super-cluster network that characterises the large-scale matter
distribution in the universe.  High-resolution simulations showing
this effect were described by the Virgo collaboration, 
\cite{colberg:virgo}. \cite{onuora:alignment} found a
significant alignment signal up to scales of $30\hMpc$ for a
$\Lambda$CDM model. \cite{faltenbacher:corr} and
\cite{hopkins_etal2004} using large scale dissipationless N-body
simulations of the $\Lambda$CDM universe report alignment of the
orientations of galaxy clusters for separations $\lesssim 100\hMpc$. 
Improved statistics due to a much larger simulation volume
enable \cite{kasun_evrard2004} to find correlations ranging up to
$200\hMpc$. \cite{basilakos_etal2005} using the same simulation than
we do, but an eight times increased mass resolution, found alignment
between cluster sized halos and their hosting super-clusters.  

This work aims to link the vector quantities of galaxy clusters
(orientation and angular momentum) with their recent mass accretion
history. The basic cluster sample comprises the 3000 most massive CDM 
halos found by a friends-of-friends (FoF) approach. Further the basic 
sample is divided into two sub-samples, namely a {\it recent 
major merger} sample and a {\it steady accretion mode} sample.
The subdivision criterion is based on the mass fraction of the two
most massive progenitors at $z=0.5$. The masses and positions of these
progenitors hold information about the pattern of infall onto the
cluster. We find a strong correlations between the infall pattern and
the cluster vector quantities. Finally, since the infall happens along
the large scale filaments we analyse long range correlations of these
vector quantities between cluster pairs by means of mark
correlation functions (see \citealt{faltenbacher:corr,kasun_evrard2004}).       

The paper is organised as follows. \S\ref{sec:sample} describes the
simulation and explains the construction of cluster sample and
sub-samples. In \S\ref{sec:impact} the impact of different mass
accretion modes (merger/steady accretion) on shape and angular
momentum of the clusters is investigated . In \S\ref{sec:mark} the
range of the alignment of the major, the minor, and the angular
momentum axes of cluster pairs is measured. \S\ref{sec:sum} summaries
the results.     
\section{Simulation and cluster sample}
\label{sec:sample}
In subsection \S\ref{sec:sim} some information about the simulation is
provided and \S\ref{sec:basicsample} describes the construction of the
basic cluster sample. In the last paragraph (\S\ref{sec:accretion})
the subdivision of the basic cluster sample into a merger and a steady
accretion mode sample is explained.      
\subsection{Simulation}
\label{sec:sim}
The simulation is performed using the public available GADGET code
(see \citealt{springel:gadget}). According to the presently favoured
cosmological model, we assume a flat universe with the
parameters $\Omega_{m}=0.3$, $\Omega_{\Lambda}=0.7$ and a Hubble
constant of $h_{100} = 0.7$. The linear power spectrum of the density
fluctuations has been normalised to $\sigma_{8}=0.9$. The simulation is 
realized within a cube of $500\hMpc$ edge length. The cold dark
matter and the gaseous component are represented by $256^3$
particles each. The mean baryonic mass fraction of the universe is
assumed to be $\Omega_{b}/\Omega_m = 0.13$, consequently the mass of a
single CDM particle is $m_{dark} = 5.4\times10^{11}\hMsol$ and the
mass of  a single gas particle is $m_{gas} =
8.1\times10^{10}\hMsol$. The gas is simulated using 
smoothed particle hydrodynamics (SPH) with no baryonic cooling or
heating processes taken into account. 

\subsection{Basic cluster sample}
\label{sec:basicsample}
For the identification of cluster halos we apply a standard
friends-of-friends (FoF) algorithm (e.g., \citealt{davis:fof}). The
CDM and the gaseous particle distributions are treated separately. The
basic cluster is based on the 3000 most massive CDM halos
found with a linking length of $b=0.17\times l$, where $l$ is the mean
particle distance (see e.g. \citealt{jenkins_etal01}). The most
massive cluster in the basic sample comprises 4201 particles which
corresponds to a CDM mass of $2.2\times10^{15}\hMsol$. The smallest 
object in the sample has a CDM mass of $1.1\times10^{14}\hMsol$
containing 209 particles. Subsequently the gaseous halos, identified
by an analogous FoF approach ($b=0.17\times l$) applied to the gas
particle distribution, are associated with their CDM clusters. 
We apply a linking length of $b=0.17\times l$ also for redshifts 
$z\neq0$.
\subsection{Mass accretion modes}
\label{sec:accretion}
\cite{suwa_etal2003} (see also \citealt{richstone_etal1992}) show that
the formation rate of galaxy clusters in the $\Lambda$CDM universe
peaks at $z\approx0.4$. In the following we want to subdivide the
basic cluster sample according to the accretion history into a 
{\it recent major merger} sample and a {\it steady accretion mode}
sample. Therefore we locate for every cluster the two most massive
progenitors at $z=0.5$, just a short time interval before the
accretion rate is maximal. These two halos at $z=0.5$
are identified as being the most massive progenitors, which contain
the highest numbers of particles also found in the current ($z=0$)
cluster. 

The basic cluster sample is subdivided based on the mass 
fraction of the two most massive CDM progenitors $m_2/m_1$, where
$m_1$ and $m_2$ denote the most and the second most massive
progenitor, respectively. The 3000 clusters are separated into three
disjoint sub-samples, each of them comprising 1000 clusters. The
recent major merger sample contains the clusters with the highest
values ($0.2\lesssim m_2/m_1\leq 
1$) for the progenitor mass fractions, meaning that the two progenitors have
comparable masses. The steady accretion mode sample contains the 1000
clusters with the lowest values ($\lesssim 0.1$) for the progenitor mass
fractions, meaning that the most massive progenitor is by far more
massive than the second most massive progenitor. Consequently all
additionally accreted mass must approach within even smaller lumps. 
The 1000 clusters with intermediate progenitor mass ratios are omitted
from any further investigation, which focuses on the impact of
the particular accretion mode. 

\begin{figure}
\epsfig{file=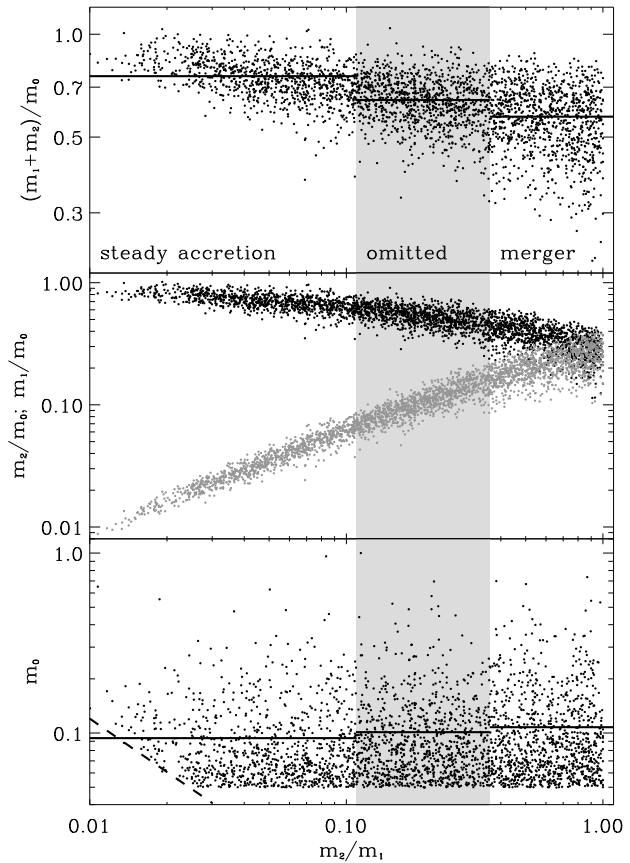,width=0.99\hsize}  
\caption{\label{fig:d3000.00.00}
The masses of the current clusters are denoted by $m_0$ and the masses of
the two most massive progenitors at $z=0.5$ are denoted with $m_1$ and
$m_2$, respectively ($m_1\ge m_2$). The upper panel shows 
the ratio of the combined mass $m_1+m_2$ divided by $m_0$ in dependence
of $m_2/m_1$. In the middle panel the ratios of $m_1/m_0$ and $m_2/m_0$
(black and gray dots, respectively) are shown. The lower panel
represents the masses of the current clusters scaled by the
mass of the largest object $\sim2.2\times10^{15}\hMsol$ at that
epoch. The shaded area entitled 'omitted'  
contains the 1000 clusters with the intermediate ratios of $m_2/m_1$,
whereas 'merger' and 'steady' indicate the sub-samples with the
highest and lowest $m_2/m_1$ respectively. The horizontal lines in the
upper and lower panels indicate the mean values of the according
sub-samples, 0.75, 0.64, 0.57 for the upper and 0.09, 0.10, 0.11 in
for lower panel. The dashed line in the lower panel marks the boundary
of completeness, below this line the second most massive progenitors
comprise less than 5 particles. 
}
\end{figure}

The upper panel of
Fig.~(\ref{fig:d3000.00.00}) shows the distribution of mass fractions
($(m_1+m_2)/m_0$) versus $m_2/m_1$ with the steady accretion mode
clusters appearing on the left and the recent major merger clusters
appearing on the right. $m_0$ denotes the mass of the emerged cluster
at $z=0$. The intermediate accretion mode clusters, which are not
investigated in the following, are located in the middle of this panel
(grey area).   
The mean values of the fractions $(m_1+m_2)/m_0$ indicated by the
horizontal lines in the upper panel of Fig.~\ref{fig:d3000.00.00}
are $\sim24\%$ lower in the merger sample compared to the steady
accretion sample, whereas $m_0$ is $\sim18\%$ higher in the merger
sample compared to the steady accretion sample. The difference in the
relative decrease of $(m_1+m_2)/m_0$ and the increase of $m_0$
indicates a sightly higher additional accretion activity in the merger
sample. The increase of $m_0$ indicates that equal mass mergers tend
to result in more massive halos. But they also show slightly enhanced 
accretion activity as indicated by the larger relative decrease of the
mean values of $(m_1+m_2)/m_0$ compared to the relative increase of
$m_0$.
\section{Impact of accretion modes on vector quantities of
clusters}
\label{sec:impact}
In this section we investigate the impact of the pattern of infall on
the principal axes and the angular momentum of current clusters. We
show the regularity of the infall of the two most
massive progenitors in the merger sample and demonstrate that an
approximate picture of the infall pattern can be derived from the
information provided by the current cluster position and the two most
massive progenitor positions (at $z=0.5$). Subsequently the
correlation of these positions with the shape and the angular momentum
of current clusters is examined. 
\subsection{Infall pattern}
\label{sec:infall}
\begin{figure}
\begin{center}
\epsfig{file=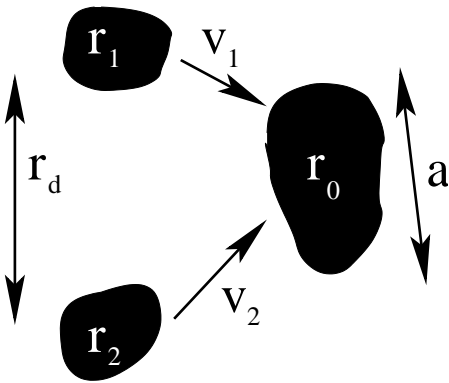,width=0.7\hsize}  
\caption{\label{fig:geometry}
The sketch depicts the vectors used to describe the geometry of the
accretion process. $\vec{r}_1$ and $\vec{r}_2$ are the locations of
the two most massive progenitors at $z=0.5$, the vector $r_d$ is the
displacement vector between theses two and $\vec{a}$ marks the orientation of
the cluster at $z=0$ (double arrows indicate that only the
direction not the sign is considered here). The vectors $\vec{v}_1$
and $\vec{v}_2$ are the velocities of the progenitors
(compare to Equ.~\ref{equ:volcities}).}
\end{center}
\end{figure}

The following illustration of the infall pattern is based
on the recent major merger sample  (see \S~\ref{sec:accretion}). The
vectors, which will be later correlated with the shape and the
angular momentum of the emerged cluster, are combinations of the
position vectors of the two most massive progenitors $\vec{r}_1$
and $\vec{r}_2$ at $z=0.5$ and of the cluster's position $\vec{r}_0$
at $z=0$. The 2000 progenitors span a mass range of $1.1\times10^{13}
\lesssim M \lesssim 4.8\times 10^{14}\hMsol$ which corresponds to a
range in particle numbers of $24 \leq N_p \leq 1062$. Subsequently we
focus on the difference vector $\vec{r}_d$ and the cross product of
the velocities $\vec{v}_\times$ as defined in the following equations.
\begin{equation}
\label{eq:dr}
\vec{r}_d = \vec{r}_1 - \vec{r}_2
\end{equation}
\begin{equation}
\label{eq:vx}
\vec{v}_\times = \vec{v}_1 \times \vec{v}_2\ ,
\end{equation} 
where 
\begin{equation}
\label{equ:volcities}
\vec{v}_i = {\vec{r}_0 - \vec{r}_i\over \Delta t}\qquad i=1,2\nonumber
\end{equation}
$\vec{r}_d$ is the difference vector of the locations of the
two most massive progenitors. The directions of the mean
velocities $\vec{v}_1$ and $\vec{v}_2$ are calculated by subtracting
the position of the current cluster by the corresponding
positions of the progenitors at $z=0.5$. Thus $\vec{v}_\times$
is perpendicular to the orbital plane of the merging
progenitors. In Fig.~\ref{fig:geometry} $\vec{r}_d$ and
$\vec{v}_\times$ are sketched along with a schematized merging event. 

\begin{figure}
\epsfig{file=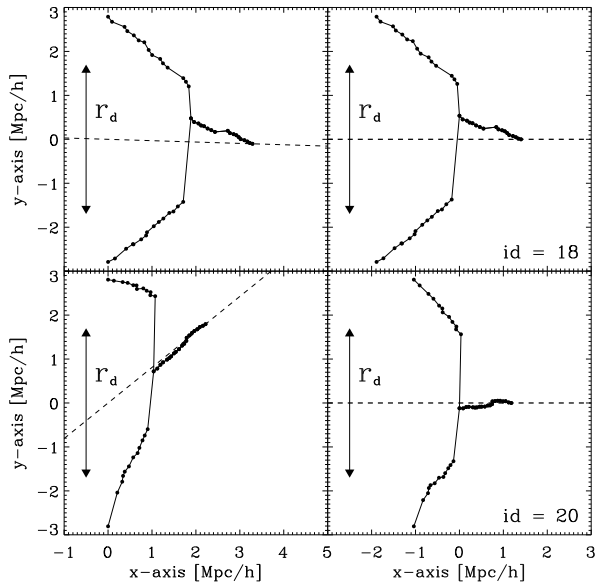,width=0.95\hsize}  
\caption{\label{fig:example}
Two examples of the adjustment procedure for the merging tracks. 
A row represents the same tracks differently stretched. The
dots indicate the comoving coordinates of the halos found in the 33
subsequent snapshots between $z=0.5$ and $z=0$. The tracks 
are projected onto the plane perpendicular to $\vec{v}_\times$ (see
Eq.~\ref{eq:vx}). The tracks on the left panels are shifted to $x=0$ 
for $z=0.5$ and equal absolute values of the y-coordinates. 
The dashed line connects the origin and the final position of the
current cluster. In the panels on the the right a correction for the
motion in y-direction is applied (i.e. shearing the coordinate system 
such that the dashed line in the left panels coincides with the
x-axis) and the merger point of the two progenitor tracks is shifted
towards the origin. These adjustments do not affect the orientation of
$\vec{r}_d$ and $\vec{v}_\times$ (see Eqs.~\ref{eq:dr} and \ref{eq:vx}).
}
\end{figure}

We only examine the tracks within the recent major merger sample to  
justify the description of the infall pattern by $\vec{r}_d$ and
$\vec{v}_\times$. Tracking the halos through 33 intermediate
snapshots in the steady accretion mode is unfeasible due to the
involvement of low mass (or sparse number of particle)
halos. Fig.~\ref{fig:example} aims to explain the projection  
and the adjustment procedures which need to be applied before all the
tracks can be stacked up. Starting at $z=0.5$ we trace the 2000 most
massive progenitors through 33 simulation outputs between $z=0.5$ and
$z=0$. By definition all of these tracks have to merge at some point
with another track. The descendant of an actual halo in the
subsequent snapshot is in the first instance associated with the halo
that contains the most overlapping particles. At the same time we
check whether the masses of the descendant and the actual halo are in
agreement. For strong deviations (factor of $\gtrsim10$) we choose, if
present, an alternate descendant showing slightly fewer overlapping
particles but better mass agreement. 

\begin{figure}
\epsfig{file=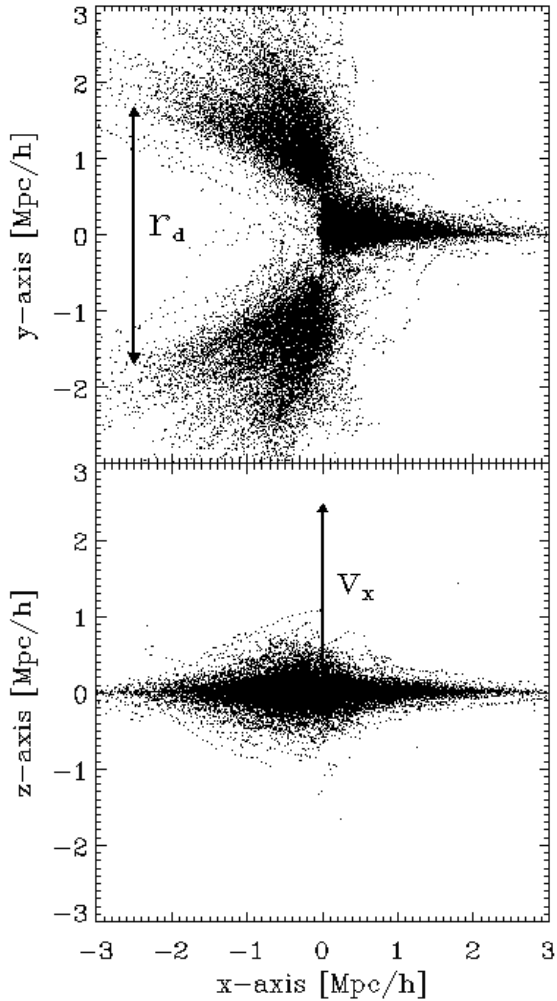,width=0.90\hsize}  
\caption{\label{fig:all}
The traces of two most massive progenitors ending in the clusters of
the merger sample are overlaid after an adjustment procedure like
sketched in Fig.(\ref{fig:example}) has been applied. 
}
\end{figure}

Fig.~\ref{fig:example} shows the projection of two different merging
events onto the {\it merging plane} perpendicular to $\vec{v}_\times$
(see Eq.~\ref{eq:vx}). Here two different representations of the same
merging event are depicted in a row. 
Each dot represents the location of the particular halo at a given
snapshot. The coordinate systems in the right panels are rotated in
such a way that the positions of the two progenitors are located on
the y-axis ($x=0$) with equal separation from the origin. The y-coordinate of
the most massive progenitor is displayed at positive y-values,
consecutively the second massive progenitor has negative y-values.
In the right panels of Fig.~\ref{fig:example}) the 
image is stretched in such a manner, that the initial coordinates
of the progenitors remain unchanged but the final cluster position 
is projected to the x-axis (i.e. the dashed line in the left
panels coincides with the x-axis in the right panels). Finally the
merger point is shifted to the origin. This adjustments neither
change the direction of $\vec{r}_d$ nor the direction of
$\vec{v}_\times$. After applying these adjustments to all the members
of the merging sample the resulting tracks are stacked up.
Fig.~\ref{fig:all} presents face and edge on sight of all the
merging events. 

As the tracks of particular merging events (Fig.~\ref{fig:example}) show
the two progenitors approach each other until they merge
in a jump-like manner. This behaviour is due to the application of the
FoF algorithm for the identification of the halos. The jumps of
$\sim1\hMpc$ to the merger point reflect roughly the radii of the
progenitor halos. The separation vector connecting the last
distinguishable positions of the progenitors is in both cases (upper
and lower panels) remarkable aligned with $\vec{r}_d$.   
This feature can also be seen in the stacked up plot
(Fig.~\ref{fig:all}). The two branches depicting the incoming
progenitor halos show enhanced densities slightly left of $x=0$. The
line joining this two areas is parallel to the y-axis. Thus
the $\vec{r}_d$ gives a good approximation of the main infall
direction. As order of magnitude estimates show the mutual
gravitational attraction of two clusters in the considered mass and
distance range dominates over the impact of large scale gravitational
field. Therefore, the two halos are expected to approach each other in
a fairly radial way, along the line joining them.

The tendency of the encounters to be parabolic can be inferred from
the average shape of the infall pattern. The flatness of the
distribution of tracks in the edge on representation (lower panel of
Fig.~\ref{fig:all}) indicates that $\vec{v}_\times$ describes the
orbital plane in a rather satisfactory manner. These outcomes confirm
that despite the apparent simplifications the vectors $\vec{r}_d$ and
$\vec{v}_\times$ are capable of determining the direction of infall
and the orbital plane, respectively.  
\subsubsection{Impact of infall pattern on clusters' shape}
\label{sec:shape}
In this paragraph we investigate the correlations between vectors
$\vec{r}_d$ and $\vec{v}_\times$ and the shapes of the current
clusters. The shape of a cluster is determined by the eigenvectors of
the second moment of mass tensor $\theta$ as given by the following
equation. 
\begin{equation}
\label{equ:second}
\theta_{ij} = \sum_{k=1,N} m_k(\vec{r}_k)\,r_{ki}r_{kj}
\end{equation}
Here $m_k$ specifies the particle masses. The position vector
$\vec{r}$  of the $N$ CDM particles belonging to the halo is given in  
respect of the centre of mass and the components of the vector
belonging to the $k$th particle are denoted $r_{ki}$ for $i=1,3$. 
Here we show only the results based on the dark matter component, 
but the behaviour of the gaseous component is very similar.
Subsequently the three eigenvectors of $\theta$ are denoted
$\vec{a}$, $\vec{b}$ and $\vec{c}$. The eigenvalues $a$, $b$ and $c$
are ordered by magnitude ($a\geq b\geq c$). Hence $\vec{a}$ indicates
the major, $\vec{b}$ the intermediate and $\vec{c}$ the minor
axis. Since cluster halos show preferentially prolate shape (see e.g.,
\citealt{warren_etal1992,faltenbacher:corr}) $\vec{a}$ is 
identified with the orientation of the cluster.  

\begin{figure}
\epsfig{file=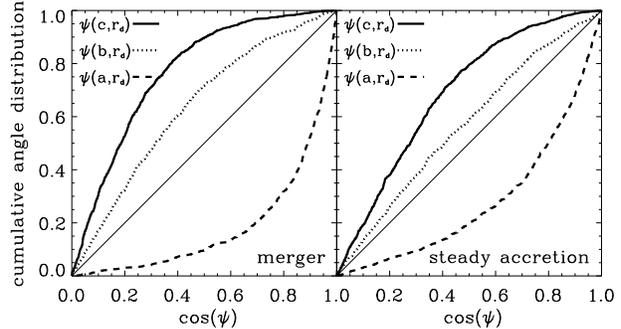,width=\hsize}  
\caption{\label{fig:rd_axes3000}
The cumulative distribution of the absolute values of the {\em cosine}
between the difference vector $\vec{r}_d$ at $z=0.5$ and the the
principal axes $\vec{a}$ (dashed), $\vec{b}$ (dotted) and $\vec{c}$
(solid) of the cluster at $z=0$ is shown. The diagonal indicates the
shape of a random distribution. The left panel gives the result from
the merger sample and the right panel shows the outcome for the steady
accretion mode sample.}  
\end{figure}

The distributions of the angles between the principal axes and
$\vec{r}_d$ are shown in Fig.~\ref{fig:rd_axes3000}. The diagonal 
represents the cumulative distribution of the {\em cosine} of the
angles between two randomly orientated 
vectors. A curve lying below the diagonal indicates an
correlation between vectors, i.e. the vectors have a tendency to be
parallel more often than a random sample.  A curve lying above the
diagonal indicates a anti-correlation between the vectors, i.e. the
vectors have a tendency to be more often perpendicular   
than in a random sample. The thick solid, dashed and
dotted lines depict the  cumulative distributions of the angles
between $\vec{r}_d$ and the principal axes $\vec{a}$, $\vec{b}$ and
$\vec{c}$ as indicated. The left panel shows the results for the
recent major merger sample and the right panel those for the steady
accretion mode sample. The dashed lines that represent the angles
between the major principal axis $\vec{a}$ and $\vec{r}_d$ deviate
most significantly from a randomly generated distribution
(diagonal). The major axis ($\vec{a}$) tends to be parallel to
$\vec{r}_d$. The result of a Kolmogorov-Smirnov (KS) test, comparing
these distributions with the random distribution, is given in
Tab.~\ref{tab:kolmogI}. 
\begin{table}
\begin{center}
\begin{tabular}{ccc}
\hline
$\psi$  & merger & steady \\\hline\hline
$(\vec{c},\vec{r}_d)$&$ \lesssim10^{-45}$&$ \lesssim10^{-45}$\\\hline
$(\vec{b},\vec{r}_d)$&$1.4\times10^{-43}$&$1.7\times10^{ -8}$\\\hline
$(\vec{a},\vec{r}_d)$&$ \lesssim10^{-45}$&$ \lesssim10^{-45}$\\\hline   
&&\\
\end{tabular}
\caption{\label{tab:kolmogI}
Results of a Kolmogorov-Smirnov test comparing the angle distributions
indicated in the first column derived from the recent major merger
sample (second column) and the steady accretion mode sample with the
random distributions as imaged in Fig.~\ref{fig:rd_axes3000}.}  
\end{center}
\end{table}
The opposite is found for the distributions associated with the two
smaller principal axes. They clearly show a trend to be
perpendicular to $\vec{r}_d$. This behaviour is expected
since the principal $\vec{a}$, $\vec{b}$ and $\vec{c}$ are
perpendicular to each other. Therefore, if a vector distribution is
strongly aligned with one of these axes it has to be perpendicular to
the other two principal axes. The different deviations 
of $\psi(\vec{b},\vec{r}_d)$ and $\psi(\vec{c},\vec{r}_d)$ from 
random expresses a strong confinement of $\vec{r}_d$ to the
$\vec{a}$-$\vec{b}$ plane (still the alignment with $\vec{a}$ is by  
far the strongest). The KS probabilities 
that the actual distributions are accidental realizations of a random
process fall below $10^{-45}$. The general trends match for the
distributions related to the two different 
samples (merger/steady). The distributions of the merger
sample show somewhat enhanced deviations from random. The shape of
clusters which underwent a recent major merging event is
obviously dominated by the orbital parameters of the two
progenitors, expressed by the vector $\vec{r}_d$. However, since the
results of the steady accretion mode sample show qualitatively equal
behaviour, the orientation of clusters must be determined in a more
general sense by the coherent mass accretion along the filaments of
the cosmic network. 

\begin{figure}
\epsfig{file=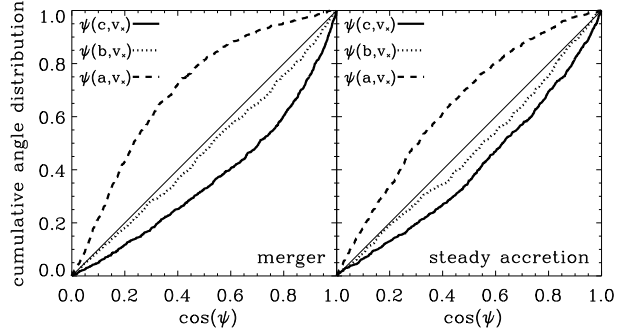,width=\hsize}
\caption{\label{fig:vxv_axes3000}
The cumulative distribution of the absolute values of the {\em cosine}
between the cross product $\vec{v}_\times$ of the two infall velocities $\vec{v}_1$ and
$\vec{v}_2$ and the the principal axes $\vec{a}$ (dashed), $\vec{b}$
(dotted) and $\vec{c}$ (solid) of the cluster at $z=0$ is imaged. The
diagonal indicates the shape of a random distribution. The left panel
gives the result from the recent major merger sample and the right panel shows the
outcome for the steady accretion mode sample.}
\end{figure}

As discussed above, \S~\ref{sec:infall}, $\vec{v}_\times$ is
perpendicular to the plane of infall and parallel to the orbital 
angular momentum of the two progenitors. In
Fig.~\ref{fig:vxv_axes3000} the solid, dotted and dashed lines
correspond to the distributions of the angles between $\vec{v}_\times$
and the minor, the intermediate and the major axis. Again the
deviation from a randomly generated angle distribution (diagonal) is
obvious. The most significant deviation from random is seen for 
$\psi(\vec{a},\vec{v}_\times)$, whereas $\psi(\vec{b},\vec{v}_\times)$
are almost compatible to random orientations. The
KS test results are displayed in Tab.~\ref{tab:kolmogII}.
The shape of the distributions indicate, that the major axis is
perpendicular to $\vec{v}_\times$, whereas the minor axes
tend to be in parallel with $\vec{v}_\times$. As mentioned above 
the distributions related to the different principal axes are not
independent. A vector distribution which is strongly aligned with one of
these axes is expected to be perpendicular to the other two axes. This
is apparently not the case for $\psi(\vec{b},\vec{v}_\times)$. This
behaviour can be explained if one takes the uncertainty in the
determination of the cluster axes into account, which accumulates in the
distribution of $\vec{b}$. On the one hand, if the cluster is prolate
(with eigenvalues $a>b\approx c$) then $\vec{b}$ and $\vec{c}$ are
degenerate, on the other hand, if the cluster is oblate ($a\approx b>c$)
then  $\vec{b}$ and $\vec{a}$ are degenerate. In total both effects
result in an almost random distribution for
$\psi(\vec{b},\vec{v}_\times)$.   
\begin{table}
\begin{center}
\begin{tabular}{ccc}
\hline
$\psi$  & merger & steady \\\hline\hline
$(\vec{c},\vec{v}_\times)$&$3.1\times10^{-40}$&$1.3\times10^{-18}$\\\hline
$(\vec{b},\vec{v}_\times)$&$1.4\times10^{ -3}$&$3.5\times10^{ -4}$\\\hline
$(\vec{a},\vec{v}_\times)$&$ \lesssim10^{-45}$&$2.3\times10^{-39}$\\\hline   
&&\\
\end{tabular}
\caption{\label{tab:kolmogII}
Results of a Kolmogorov-Smirnov test comparing the angle distributions
indicated in the first column derived from the recent major merger sample (second
column) and the steady accretion mode sample with the random
distributions as plotted in Fig.\ref{fig:vxv_axes3000}.}  
\end{center}
\end{table}
The distributions associated with the merger events show a more
prominent deviation from the random  for
$\psi(\vec{a},\vec{v}_\times)$ and $\psi(\vec{c},\vec{v}_\times)$ 
and a slightly more random like distribution for
$\psi(\vec{b},\vec{v}_\times)$ compared to the steady accretion mode
sample. However (as seen before in the $\vec{r}_d$ distributions)
the similarity of the $\vec{v}_\times$ distributions
indicates that the mass accretion along the filaments is the 
predominant factor in the determination of the orientation, no matter 
which accretion mode (merger/steady) is considered.  
\subsubsection{Impact of infall pattern on clusters' angular momentum}
\label{sec:angular}
\begin{figure}
\epsfig{file=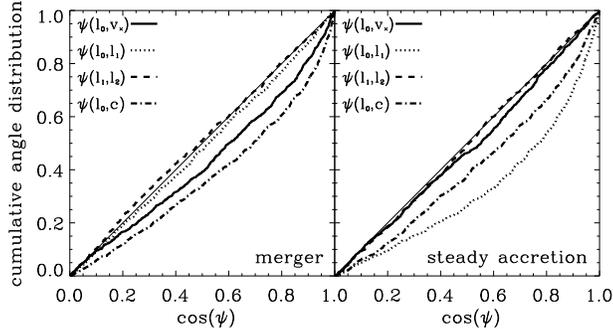,width=\hsize}  
\caption{\label{fig:vxv_ang3000}
The cumulative distribution of the absolute values of the {\em cosine}
between; 1. the angular momentum vector $\vec{l}_0$ at $z=0$ and the cross
product $\vec{v}_\times$ of the two infall velocities $\vec{v}_1$ and 
$\vec{v}_2$ at $z=0.5$  (solid line), 2. $\vec{l}_0$ and the minor principal
axis $\vec{c}$ (dashed-dotted), 3. $\vec{l}_0$ and the angular momentum
vector $\vec{l}_1$ of the most massive progenitor at $z=0.5$ (dotted
line), 4. $\vec{l}_1$ and the the angular momentum vector $\vec{l}_2$
of the second most massive progenitor at $z=0.5$ (dashed line). The
diagonal indicates the shape of a random distribution. The left panel
gives the result from the recent major merger sample and the right
panel shows the outcome for the steady accretion mode sample.}
\end{figure}

The angular momentum is closely related to the dynamics of
the halo. However, since the spin parameter $\lambda$ of halos is
roughly $0.05$ the ordered rotational support against gravity is only
about 5\% (\citealt{steinmetz_bartelmann1995,cole_lacey1996}). Thus
the correlation analysis using the direction of the angular momentum
is expected to be blurred by this effect. The angular momentum for a
FoF cluster is computed in the following way.   
\begin{equation}
\vec{l} =  \sum_{i=1,N}m_k\ \vec{r}_i\times\vec{v}_i
\end{equation}
The $\vec{r}$ denotes the particle positions relative to centre of
mass, $m_k$ is the particle mass and $N$ is the number of particles
belonging to the halo. As discussed above the vector $\vec{v}_\times$
is perpendicularly orientated to the infall plane. If the angular
momentum $\vec{l}_0$ of the emerged cluster at $z=0$ is dominated by
the orbital angular momentum of two most massive progenitors, then a
correlation between $\vec{v}_\times$ and  
$\vec{l}_0$ is expected. Therefore we analyse subsequently 
the distribution of the angles between $\vec{l}_0$ and
$\vec{v}_\times$. Additionally we consider the angular momenta of 
the two progenitor halos $\vec{l}_1$ and $\vec{l}_2$, where $\vec{l}_1$
corresponds to the angular momentum of the most massive and
$\vec{l}_2$ to the second most massive progenitor. 
In Fig.~\ref{fig:vxv_ang3000} the various angle distributions
connected to the different angular momenta are displayed. Since we
have found a strong correlation between $\vec{v}_\times$ and $\vec{c}$
(minor axis of the cluster) we also show $\psi(\vec{l}_0,\vec{c})$
using dashed-dotted lines. On the left hand side the results for the
merger sample are depicted and the outcome for the steady accretion
mode sample is displayed on the right panel. The according KS test
results comparing these distributions with the random distributions
(diagonals in Fig.~\ref{fig:vxv_ang3000}) are presented in
Tab.~\ref{tab:kolmogIII}.  
\begin{table}
\begin{center}
\begin{tabular}{ccc}
\hline
$\psi$  & merger & steady \\\hline\hline
$(\vec{v}_\times,\vec{l}_0)$&$2.1\times10^{-13}$&$8.1\times10^{ -3}$\\\hline
$(\vec{l}_0,\vec{l}_1)$     &$7.4\times10^{ -2}$&$ \lesssim10^{-45}$\\\hline
$(\vec{l}_1,\vec{l}_2)$     &$3.4\times10^{ -1}$&$7.5\times10^{ -1}$\\\hline   
$(\vec{l}_0,\vec{c})$       &$1.0\times10^{-35}$&$2.3\times10^{-17}$\\\hline
&&\\
\end{tabular}
\caption{\label{tab:kolmogIII}
Results of a Kolmogorov-Smirnov test comparing the angle distributions
indicated in the first column derived from the recent major merger
sample (second column) and the steady accretion mode sample with the
random distributions as presented in Fig.\ref{fig:vxv_ang3000}.}  
\end{center}
\end{table}
The distributions between the angular momenta of the two progenitors
$\psi(\vec{l}_1,\vec{l}_2)$ agree with random orientations for both
samples. $\psi(\vec{l}_0,\vec{c})$ also shows similar behaviour,
indicating that $\vec{l}_0$ tends to be parallel to
$\vec{c}$. Here the KS probabilities are enhanced for the merger
sample, but the qualitative trend is similar in the steady accretion
mode sample. The most obvious differences between the two samples 
are seen in $\psi(\vec{l}_0,\vec{v}_\times)$ and
$\psi(\vec{l}_0,\vec{l}_1)$. These distributions show opposite 
behaviour for the merger and the steady accretion mode sample. 
In the merger sample the KS probabilities for
$\psi(\vec{l}_0,\vec{v}_\times)$ are very low, indicating that this
distribution is obviously incompatible with a random distribution.
Instead $\psi(\vec{l}_0,\vec{l}_1)$ is almost in agreement
with a random distribution. The opposite picture arises in the steady
accretion mode sample. Here $\psi(\vec{l}_0,\vec{l}_1)$ is clearly not
in agreement with a random distribution, whereas
$\psi(\vec{l}_0,\vec{v}_\times)$ can't be differentiated from random.

We find the effect of steady accretion on the angular momentum
is in line with previous work (i.e.~\citealt{vitvitska:origin}).
If there is no major merger, the angular momentum of the most massive 
progenitor will be aligned with the angular momentum of the 
descendant cluster.  As seen in Fig.~\ref{fig:d3000.00.00}, the 
most massive progenitor in the steady accretion sample is orders
of magnitude larger than all other infalling halos.  This explains the
result for $\psi(\vec{l}_0,\vec{l}_1)$. The small tendency of an
alignment between $\vec{l}_0$ and $\vec{v}_\times$ may have the same
origin as the finding by \cite{benson2005}, that merging dark matter 
satellites show a weak tendency to have velocities perpendicularly
orientated to the angular momentum of the main halo. 

In the merger sample the orbital angular momentum of the two most
massive progenitors shows a strong impact on the angular momentum of
the descendant. A decided difference between
$\psi(\vec{l}_0,\vec{v}_\times)$ and a random distribution can be
seen. The KS probability of $2.1\times10^{-13}$ strongly discounts
any random process as being the origin of this distribution. The tendency 
for $\vec{l}_0$ and $\vec{v}_\times$ to be parallel confirms
the picture that a large fraction of the descendant's angular momentum  
was gained from the orbital angular momentum of the two most massive
progenitors. This result is in agreement with the detailed study of
the angular momenta of dark matter halos by \cite{vitvitska:origin}.  
Whereas the correlation of $\vec{l}_1$ and $\vec{l}_0$
is in agreement with random orientations as can be seen by the rather 
high value of the KS test. 
\section{Correlations in orientation and angular momentum}
\label{sec:mark}
As shown above orientation and angular momentum of current
clusters are tightly connected to the infall pattern. The dominant
amount of mass is accreted along filaments. Therefore we employ the
technique of {\it mark correlation functions} (see
\cite{stoyan:fractals,faltenbacher:corr,kasun_evrard2004}) to 
investigate the difference in the large scale correlations between the
basic sample and the two accretion mode sub-samples
(merger/steady). \cite{faltenbacher:corr} used the following equation
to study {\it filamentary} alignment.
\begin{equation}
\label{equ:def-vector-corr}
\CF_{\vec{v}}(r) = \frac{1}{2}
\paverage{|\vec{v}_1\cdot{\hat{\vec{r}}}| + |\vec{v}_2\cdot{\hat{\vec{r}}}|}(r)\ , 
\end{equation}
Here $\vec{v}$ is a normalised vector quantity ($|\vec{v}|=1$),
$\vec{r}$ is the distance vector between a pair of clusters with
absolute value $r$. The normalised distance vector is denoted by 
$\hat{\vec{r}}=\vec{r}/r$ and $\left\langle\ \right\rangle_{\rm P}$
denotes the average over all pairs with separation $r$. 
Due to the spatial symmetry the
absolute values of these scalar products are used. $\CF(r)$ quantifies
the $\CF$ilamentary alignment of the vectors $\vec{v}_1$ and
$\vec{v}_2$ with the line connecting both clusters. $\CF(r)$ is
proportional to the mean value of the {\em cosine} of the angles
between $\vec{v}_1$ and the vector $\hat{\vec{r}}$ connecting the
pairs. Random orientations imply $\CF(r)=0.5$.
Since the distribution of angles is not of Gaussian shape around 
a given mean value, it would be wrong to apply Gaussian statistics
to estimate the the errors. Thus, to investigate the 1$\sigma$
confidence intervals we perform a non-parametric Monte Carlo test
\citep{besag:stat,gottloeber:spatial,kerscher:gauss,kerscher:reg}.
Here the null hypothesis is ``no deviation from random
angles''. We simulate this null hypotheses by keeping the  
positions of the halos fixed and randomly re-assigning the orientations
to the halos. This procedure is repeated 99 times. The rms-deviation
for these realisations gives a measure for the 1$\sigma$ confidence
interval for the given data set. These intervals are indicated by the
shaded areas around the expected values for random distributions in
Fig.~\ref{fig:link054.0.170}. Thus, if $\CF(r)$ lies outside of this 
intrinsic scatter region around $0.5$ the deviation from random
orientations is confirmed at a 1$\sigma$ confidence level. 
In addition to the alignment of the CDM component we
present the alignment signals for the gaseous
halos. \cite{kazantzidis_etal2004} (see also \citealt{dubinski1994})
pointed out that the elongation of halos is reduced due to radiative
cooling and subsequent condensation of baryonic matter at the cluster
centre. Since we use non radiative SPH simulations this effect is not 
represented. However 
according to Eq.~\ref{equ:second} the second moment of mass tensor
$\theta$ and thus the orientations of the principal axes are dominated
by the mass distribution on the outskirts of the clusters. Moreover,
we are mainly concerned with the orientation of the halo, not the
actual axial ratios, and the orientation is not affected by
dissipative effects. 
\subsection{Correlation of clusters' orientations}
\begin{figure*}
\epsfig{file=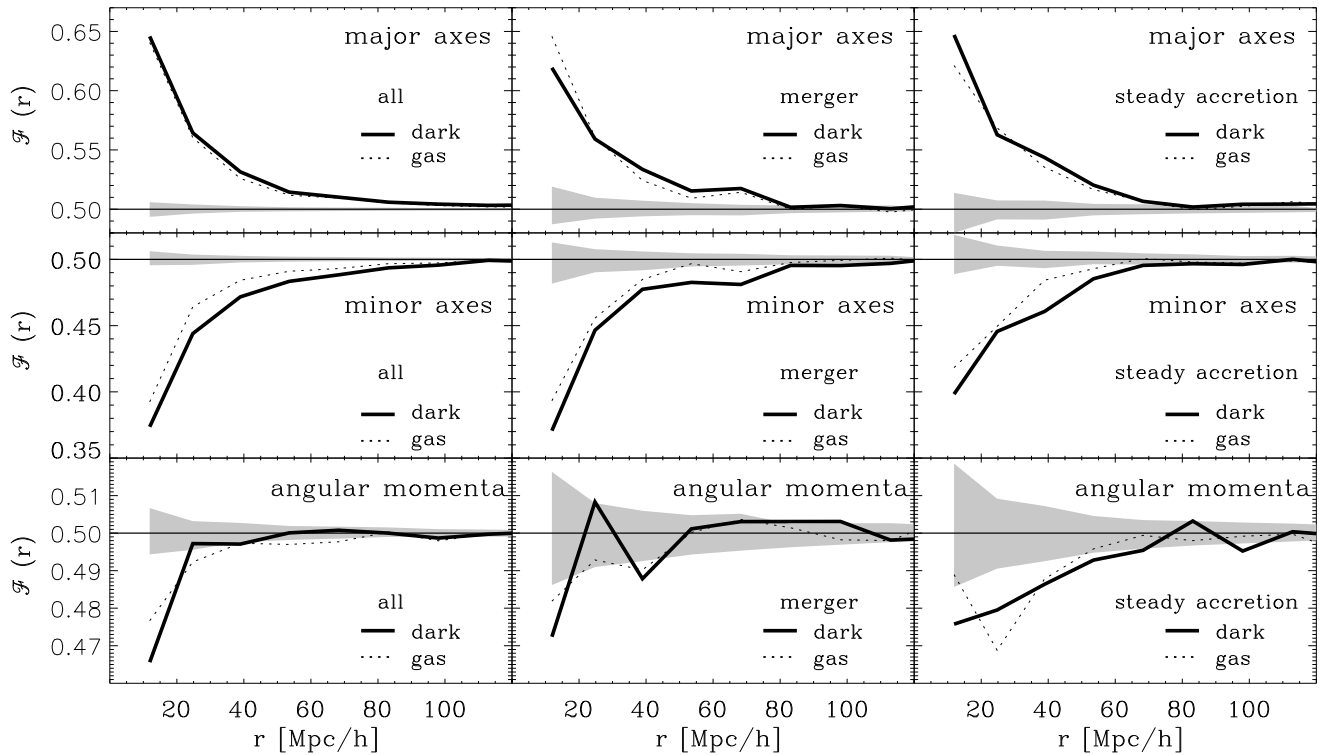,width=\hsize}  
\caption{\label{fig:link054.0.170}
The upper panel shows the mark correlation function, using the
clusters' major axes as vector marks. The solid lines display the
signal for the CDM component. The dotted lines show the behaviour of
the gas. The horizontal solid line gives the expected values 
for a random distribution of the orientations. The shaded area is a
measure of the intrinsic scatter of the signal. The three
panels from right to left image the results for the whole, the merger
and the steady accretion sample, respectively. In the middle panels
the correlation for the minor axes of the clusters are shown and the
lower panels represent the signal obtained for the angular momentum.} 
\end{figure*}

We have shown in \label{sec:shape} that the different accretion modes
do not result in a qualitatively different impact on the shape of current
cluster. However the correlation between shapes and infall pattern are
reduced in the steady accretion mode sample. Here we separately
investigate the filamentary correlation of the orientation of cluster
for the different samples (basic, merger, steady) to see whether this
quantitative differences influence the signal for large scale
correlations. Replacing $\vec{v}$ in 
Eq.~\ref{equ:def-vector-corr} with the major (and subsequently the
minor principal) axes results in 
\begin{equation}
\label{equ:mcfori}
\CF_{\vec{p}}(r) = \frac{1}{2}
\paverage{|\vec{p}_1\cdot{\hat{\vec{r}}}| +
|\vec{p}_2\cdot{\hat{\vec{r}}}|}(r)\ .
\end{equation}
Here $\vec{p}_1$ and $\vec{p}_2$ denote the principal axes (major or
minor) of a cluster pair at $z=0$ and the associated separation vector
$\vec{r}$. The analysis is done for the whole cluster sample and for
the two sub-samples (merger/steady) separately. The results for the
alignment of the major and the minor axes are shown in 
upper and middle panel of Fig.~\ref{fig:link054.0.170}.

$\CF(r) > 0.5$ indicates that the axes tend to be aligned with the
connecting line of the clusters. For the basic cluster sample,
comprising the 3000 most massive clusters in the simulation volume,
this signal for the major axes extends up to $100\hMpc$ (left panel of
Fig.~\ref{fig:link054.0.170}, compare also to Fig.~3b in
\citealt{faltenbacher:corr}). The signal for the merger and the  
steady accretion mode samples extends up to $80\hMpc$. There is no
distinctive difference between the results for the two sub-samples. 
The somewhat shorter range seen in both sub-samples appears to be a
result of the lower statistics (only 1000 members per sub-sample)
rather than caused by the different accretion modes.

$\CF(r) < 0.5$ indicates that the axes tend to be perpendicular
to the connecting line of the cluster pairs. In the basic cluster
sample the signal for the minor axis extends up to $100\hMpc$ (middle
left panel of Fig.~\ref{fig:link054.0.170}). This result is in good
agreement with \cite{bailin_steinmetz2004}, where a strong tendency
for the minor axes of halos to lie perpendicular to the large scale
filaments is found. The CDM components of the recent major merger
sample (middle panel of Fig.~\ref{fig:link054.0.170}) and the steady
accretion mode sample show a tendency to be perpendicular to the
connecting line for separations up to $70\hMpc$. Here again the
shorter correlation ranges for the sub-samples are likely be caused by
the smaller sample size. 

The comparison between the signals for the CDM component (thick solid
line) to the gaseous component (thin dotted line) generally show good
agreement. However, the small differences in the correlation range are
likely to mirror the different physical properties of the two
components. Using X-ray position angles \citealt{chambers:nearest}
found that clusters with separations $d_n\leq20\hMpc$ tend to point
towards their neighbours. Their sample comprised 45 clusters thus the 
statistics are fairly low compared to our analysis.

Previously we have shown that the infall of halos is correlated  
with the orientation of the cluster and is independent of accretion mode.
In this section we demonstrate that the orientations of clusters are 
correlated and independent of accretion mode.  Since the steady accretion
mode sample shows just as much correlation as the merger mode sample all 
progenitors must be infalling along the same relative direction.  From
this, we conclude that the alignment is strongly supported by infall
along filamentary structure.  The correlation length ($\lesssim 100\hMpc$) 
for cluster alignment that we find in our simulation is in agreement with
the findings of \cite{hopkins_etal2004}.
\subsection{Alignment of clusters' angular momenta}
In \S~\ref{sec:angular} we have analysed the effects of different
accretion modes on the angular momentum of current clusters. We find a
different behaviour in the merger and the steady accretion mode
samples. To see, whether these differences have impact on the large
scale correlations,  we investigate the correlation of the intrinsic
angular momenta of cluster pairs in the whole sample, as well as in
the two sub-samples. Therefore the angular momenta is used as vector
mark in equation (\ref{equ:def-vector-corr}).     
\begin{equation}
\label{equ:mcfang}
\CF_{\vec{l}}(r) = \frac{1}{2}
\paverage{|\vec{l}_a\cdot{\hat{\vec{r}}}| + |\vec{l}_b\cdot{\hat{\vec{r}}}|}(r) 
\end{equation}
Here $\vec{l}_a$ and $\vec{l}_b$ denote the normalised angular momenta
of a cluster pair separated by the spatial vector $\vec{r}$. The result
of this study is displayed in the lower panels of
Fig.~\ref{fig:link054.0.170}. Similar to the result for the 
correlation of the minor axes (middle panels
Fig.~\ref{fig:link054.0.170}) the angular momenta of the CDM component
show the trend to be perpendicular to the connecting lines. Thus they
tend to be perpendicular to the filaments as well. The signal
reaches only out to separations of $\sim20\hMpc$ for the complete
sample. This result is in qualitative agreement with Fig.~7b in 
\cite{faltenbacher:corr}. 

The CDM signal in the recent major merger sample is hardly 
distinguishable from random fluctuations. Furthermore the deviation
from random is far smaller than seen in the correlations of the principal
axes. As mentioned in the beginning of \S\ref{sec:angular}, coherent
rotation supports the halo against gravitational collapse only at a
5\% level, thus a low deviation from random may be expected. However,
notable is the long correlation range of the steady accretion mode
sample ($\sim60\hMpc$) compared to the almost disappearing signal for  
the merger sample. According to the established picture of halos'
mass growth they go through a merger phase (rapid mass growth)
prior to a steady accretion phase (see
e.g. \citealt{wechsler_etal2002}). Therefore, on average the merger  
sample is younger than the steady accretion sample. The results 
from the above combined with those from the previous section lead to the
conclusion that infall along the filaments tends to torque
the halo's angular momentum  into a perpendicular orientation with
respect to those filaments. \cite{colberg_etal2005} finds a trend
that more massive clusters have more associated
filaments. Fig.~\ref{fig:d3000.00.00} indicates a relative  
increase of clusters' masses in the major merger sample. Moreover, the
mass accretion in addition the two most massive progenitors is maximal
in the major merger sample. The rapid accretion in the merger sample
takes place along various filaments which causes violent changes in the
the angular momentum. In a later state, when the smaller filaments are
exhausted, the infall direction is more uniform. According to this
scenario it is expected that the merger sample will settle down in
time and reflect the same sort of correlations as the current steady
accretion sample.  We plan on addressing this issue in a soon to be
forthcoming paper \cite{allgood_etal2005}.

The slightly different behaviour of the gaseous component
indicated by the thin dotted lines in the lower panel of
Fig.~\ref{fig:link054.0.170} is likely to be caused by differences in 
the underlying dynamics. The gaseous component mainly deviates for 
cluster pair separations $\lesssim20\hMpc$ from the dark matter
distribution.
\section{Summary}
\label{sec:sum}
We use large scale SPH+N-body simulations within the
concordance $\Lambda$CDM model to measure the mass accretion histories
and the correlation of orientation and angular momentum of galaxy
clusters. The two most massive progenitors at $z=0.5$ are located 
for the 3000 most massive current clusters. Based on the mass
fractions of these two progenitors $m_2/m_1$ ($m_1\geq m_2$) two
sub-samples were constructed, a recent major merger sample
containing the 1000 clusters with highest $m_2/m_1$ values and a
steady accretion mode sample with the 1000 lowest $m_2/m_1$ values. 

The mean values of $(m_1+m_2)/m_0$, where $m_0$ denotes the current mass
of the particular cluster, for the recent major
merger sample and the steady accretion mode sample are $0.57$ and
$0.75$, respectively. The clusters which experienced recent major
mergers tend to accrete even more matter in addition to $m_1+m_2$ 
than silently evolving clusters. Major mergers predominantly happen to
clusters showing averagely enhanced mass accretion. Moreover these 
clusters tend to have higher masses than their non merging counterparts. 
This outcome can be related to the work by
\cite{colberg_etal2005}. They demonstrate that more massive clusters
tend to have more filaments approaching them. Consequently their
over-all accretion activity may be enhanced. Also \cite{benson2005}
reports strongly enhanced merging activities for the largest members
in his sample of N-body clusters.   

The reconstruction of the halo tracks in the merger sample reveals
that the infall pattern are not complex. The merging takes place
along the vector $\vec{r}_d$ joining the two most massive progenitor
halos at $z=0.5$. Despite the relatively high additional mass
accretion the tracks of the two most massive progenitors are
reasonably well confined within the plane defined by $\vec{v}_\times$
(Eq.~\ref{eq:vx}). The tracks do not twine each other, instead 
the two most massive progenitors approach each other parallel to the 
line joining them $z=0.5$. This behaviour is expected since the mutual
gravitational attraction of two clusters in the considered mass and
distance range dominates over the impact of large scale gravitational
field.

We find that the orientation and the angular momentum of
current clusters are tightly correlated with the vectors $\vec{r}_d$
and $\vec{v}_\times$. $\vec{r}_d$ tends to be parallel to the
major axis and perpendicular to the minor axis of the cluster.
The opposite behaviour is found for $\vec{v}_\times$. Both accretion
mode sub-samples show similar correlations. This indicates that
the current orientation is a result of mass flows along the filaments onto
the cluster, independent of accretion mode. It is likely that the
accretion flow determines the velocity structure of the cluster. If
this is the case our results agree with alignment of position and
velocity principal axes found by \cite{tormen1997} and
\cite{kasun_evrard2004}. However, the angular momentum behaves
differently in the two sub-samples. The angular 
momenta $\vec{l}_0$ of the clusters in the merger sample show the
tendency to be parallel to $\vec{v}_\times$, whereas the orientation
of the intrinsic angular momenta of the most massive progenitors
$\vec{l}_1$ at $z=0.5$ has a minor influence on the angular momentum
at $z=0$. The opposite picture arises in the steady accretion mode
sample. Here $\vec{l}_0$ and $\vec{l}_1$ are correlated, whereas the
angle distribution between $\vec{l}_0$ and $\vec{v}_\times$ is in
agreement with random orientations. 

The major axes of cluster pairs in the basic sample tend to
be parallel to their connecting line for distances up to
$\sim100\hMpc$. Correlations up to $\sim80\hMpc$ are found in the
steady accretion and the merger sample. The reduced extension of the
correlation is likely caused by the reduced statistic in the
sub-samples. The minor axes of cluster pairs tend to be
perpendicular to their connecting line. The extension of these
correlations is similar to the findings for the major axes. 
The appearance of large scale correlations of cluster orientations has
also been shown by  \cite{faltenbacher:corr,hopkins_etal2004} and
\cite{kasun_evrard2004}. The mass accretion along the filaments
appears to be responsible for the signal, since the accretion
mode shows no impact on the correlation signal.  

Weak correlations are found for the angular momenta in the complete and
the steady accretion mode sample. The angular momenta tend to be
orthogonal to the connecting lines of cluster pairs up to $\sim20\hMpc$
for the complete and up to $\sim60\hMpc$ for the steady accretion mode
sample. The orthogonality between angular momentum and filaments is 
a result of mass accretion along these filaments (see also
\citealt{bailin_steinmetz2004}). The signal is stronger for the steady
accretion sample because the halos in this sample have already gained
major parts of their angular momentum from infall along the filaments.
The merger sample is still in a phase of accretion and it is expected
that additional infall along the filaments will bring the merger
sample into better agreement with the steady accretion sample.  The
correlations in the orientations and the angular momenta for the gas
and the dark matter components show very similar behaviour, supporting
the search for alignment effects in large X-ray samples of clusters of
galaxies. 
\section*{Acknowledgements}
The authors would like to thank the anonymous referee for helpful and
constructive comments. We would also like to thank William G. Mathews
and Joel Primack for insightful advises. AF has been supported by 
NSF grant (AST 00-98351) and NASA grant (NAG5-13275). BA has been supported
by NASA grant (NAG5-12326) and a NSF grant (AST-0205944).  GY thanks
M.E.C (Spain) for financial support under projects  AYA-2003-07468 and
BFM2003-01266. GY and SG thanks the Acciones Integradas
Hispano-Alemanas for support. YH has been supported by  the Israel
Science Foundation grant 143/02. The simulations were performed at the
John von Neumann Institute for Computing J\"ulich.

\end{document}